\documentstyle[12pt,psfig]{article}

\newcommand{\be}{\begin{equation}}
\newcommand{\ee}{\end{equation}}

\begin{document}

\begin{center}

{\bf EQUATION OF STATE IN QUANTUM CHROMODYNAMICS} \\ [5mm]

V. I. Yukalov$^{1\dag}$ and E. P. Yukalova$^2$ \\ [5mm]

{\small

(1) {\it Bogolubov Laboratory of Theoretical Physics \\
Joint Institute for Nuclear Research, Dubna 141980, Russia} \\
(2) {\it Laboratory of Computing Techniques and Automation \\
Joint Institute for Nuclear Research, Dubna 141980, Russia} \\

$\dag$ {\it E--mail: yukalov@thsun1.jinr.ru} }

\end{center}

\vskip 5mm

\begin{center}
\begin{minipage}{150mm}
\centerline{\bf Abstract}

The equation of state for quark--gluon plasma is obtained, being valid 
for arbitrary values of the coupling parameter and of temperature. This 
equation is constructed from perturbative expansions for free energy and 
for renormalization function. Summation of asymptotic perturbative series 
is accomplished by means of the self--similar approximation theory. \\


\end{minipage}
\end{center}

\vskip 10mm

\section{Introduction}

Under an equation of state one implies the dependence of pressure or free 
energy on some characteristic parameters, for instance, coupling 
parameter or temperature. The definition of pressure is
$$
p = \frac{T}{V} \ln\; {\rm Tr}\; e^{-H/T} \; ,
$$
where $T$ is temperature; $V$, volume; and $H$ is the Hamiltonian 
defining the considered system. In recent years there has been much 
attention paid to understanding the equation of state in quantum 
chromodynamics. The state of matter, called quark--gluon plasma, almost 
certainly existed in the early Universe up to $10^{-5}$s after the Big 
Bang and is likely to be found in the interior of neutron stars. One 
hopes that this state can be formed in the collisions of relativistic 
heavy nuclei where it can be studied in a controlled and systematic way 
[1--6]. The formation of quark droplets in colliding nuclei can occur at 
already existing accelerator energies, leading to the cumulative effect [7].

There are two ways of obtaining the equation of state in quantum 
chromodynamics. One is a numerical way based on lattice simulation, and 
another way is by employing statistical models. A discussion of these 
ways can be found in the recent review [6]. In this report we advance the 
third possibility of deriving the equation of state in quantum 
chromodynamics. This novel way is based on the summation of perturbative 
series in powers of the coupling parameter. The method of summation we 
use here is a variant of the self--similar approximation theory [8--15] 
employing self--similar exponential approximants [16]. This method has 
been successfully used for constructing accurate equations of state for 
several physical systems [15,16].

\section{Summation of series}

Perturbation theory can be used in {\it QCD} at high temperatures when 
the coupling parameter becomes small. One usually considers the case of 
zero chemical potential and assumes that the temperature is high enough 
that fermion masses can be ignored. An expansion for the {\it QCD} free 
energy or pressure, in powers of the coupling parameter $g$ has been 
known to fourth order [17--20] and recently was calculated to fifth order 
[21,22].

The perturbative expansion for the pressure of high--temperature gauge 
theory with massless fermions in four dimensions can be written [21] as
\be
p(g) \simeq \frac{\pi^2d_A}{45}\; T^4 \sum_n ( c_n + c_n'\ln g ) g^n \; ,
\ee
where the coefficient are
$$
c_0 = 1 + \frac{7d_F}{4d_A} \; , \quad c_1=0 \; , \quad
c_2 = - \frac{5}{(4\pi)^2}\left ( C_A +\frac{5}{2} S_F\right ) \; , \quad
c_3 = \frac{240}{(4\pi)^3}\left (\frac{C_A + S_F}{3}\right )^{3/2} ,
$$
$$
c_4 = -\frac{5}{(4\pi)^4}\left [ C_A^2\left ( 
\frac{38\zeta'(-3)}{3\zeta(-3)} - \frac{148\zeta'(-1)}{3\zeta(-1)} +
\frac{64}{5} - 4\gamma_E + \frac{22}{3} \ln\frac{\mu}{4\pi T}\right ) +
\right.
$$
$$
+ C_A S_F\left ( \frac{\zeta'(-3)}{3\zeta(-3)} - 
\frac{74\zeta'(-1)}{3\zeta(-1)} + \frac{1759}{60} + \frac{37}{5}\ln 2 -
8\gamma_E + \frac{47}{3}\ln\frac{\mu}{4\pi T}\right ) + $$
$$
+ S_F^2\left ( \frac{8\zeta'(-3)}{3\zeta(-3)} - 
\frac{16\zeta'(-1)}{3\zeta(-1)} - \frac{1}{3} + \frac{88}{5}\ln 2 -
4\gamma_E - \frac{20}{3}\ln\frac{\mu}{4\pi T}\right ) + 
$$
$$
\left. + S_{2F}\left ( 24\ln 2 - \frac{105}{4}\right ) -
48 C_A( C_A + S_F ) \ln\left ( \frac{1}{2\pi}\sqrt{\frac{C_A + S_F}{3}}
\right )\right ] \; ,
$$
$$
c_5 = \frac{5}{(4\pi)^5}\sqrt{\frac{C_A + S_F}{3}} \left [ C_A^2 \left (
264\ln 2 - 494 - 24\pi^2 + 176\gamma_E + 176\ln\frac{\mu}{4\pi T}\right ) +
\right.
$$
$$
+  C_A S_F\left ( 72 - 128\ln 2 + 112\gamma_E + 
112\ln\frac{\mu}{4\pi T} \right ) + 
$$
$$
\left. + S_F^2\left ( 32 - 128\ln 2 -64\gamma_E
- 64 \ln\frac{\mu}{4\pi T}\right ) - 144 S_{2F}\right ] \; ,
$$
$$
c_0'=c_1'=c_2'=c_3'=c_5'=0\; , \qquad
c_4' =\frac{240}{(4\pi)^4}\; C_A (C_A + S_F ) \; .
$$
Here the dimensionless regularization is used, the scale $\mu$ 
corresponds to the modified minimal subtraction scheme, $\zeta(\cdot)$ 
is the Riemann zeta function, $\gamma_E$ is the Euler--Mascheroni constant,
and for $SU(N_c)$ theory, with $N_c$ colours and with $n_f$ fermions in 
the fundamental representation,
$$
d_A = N_c^2  - 1 \; , \qquad d_F = N_c\; n_f \; , \qquad C_A = N_c \; , 
$$
$$
S_F = \frac{1}{2}\; n_f \; , \qquad S_{2F} = \frac{N_c^2 -1}{4N_c}\; n_f \; .
$$
For $QCD$ with $N_c=3$, one has
$$
d_A=8 \; \qquad d_F=3n_f \; , \qquad C_A=3 \; , \qquad 
S_F = \frac{1}{2}\; n_f \; , \qquad S_{2F} = \frac{2}{3}\; n_f \; .
$$
Then expansion (1) can be written as
\be
p(g) \simeq \frac{8\pi^2}{45} T^4 \left ( \sum_n c_n g^n +
\ln g\sum_n c_n' g^n\right ) \; ,
\ee
with the nonzero coefficients
$$
c_0 = 1 +\frac{21}{32} n_f \; , \qquad c_2 = - 0.09499\left ( 1 +
\frac{5}{12} n_f\right ) \; , \qquad 
c_3 = 0.12094 \left ( 1 +\frac{1}{6} n_f \right )^{3/2} ,
$$
$$
c_4 = 0.04331\left ( 1 + \frac{1}{6} n_f\right )
\ln\left ( 1 + \frac{1}{6} n_f\right ) +
$$
$$
+ 0.01733 - 0.00763 n_f - 0.00088 n_f^2 - 
0.01323\left ( 1 +\frac{5}{12}n_f \right )
\left ( 1 - \frac{2}{33} n_f\right ) \ln\frac{\mu}{T} \; ,
$$
$$
c_5 = - \left ( 1 +\frac{1}{6} n_f\right )^{1/2} \left (
0.12806 + 0.00717 n_f - 0.00027 n_f^2 \right ) +
$$
$$
+ 0.02527\left ( 1 +\frac{1}{6} n_f \right )^{3/2}\left ( 
1 - \frac{2}{33} n_f\right )\ln\frac{\mu}{T} \; , 
$$
$$
c_4' = 0.08662 \left ( 1 + \frac{1}{6} n_f \right ) \; .
$$

It is convenient to introduce the dimensionless function
\be
f(g) \equiv \frac{p(g)}{p(0)} \; , \qquad
p(0) = \frac{8\pi^2}{45} \left ( 1 + \frac{21}{32} n_f \right ) T^4 \; ,
\ee
being pressure (2) reduced to the Stefan--Boltzmann limit $p(0)$. As
$g\rightarrow 0$, the reduced pressure (3) reads
\be
f(g) \simeq \sum_{n=0} a_n g^n + \ln g \sum_{n=2} a_{2n}'g^{2n} ,
\ee
where $a_n\equiv c_n/c_0$ and $a_n'\equiv c_n'/c_0$. Truncating the 
asymptotic series (4), we have the sequence of perturbative approximations
\be
f_k(g) = 1 + a_2g^2 + \ldots + a_k g^k \; ,
\ee
for $k=2,3,4,5$, and
\be
f_6(g) = 1 + a_2 g^2 + \ldots + a_6 g^6 + \ln g \left (
a_4' g^4 + a_6' g^6\right )
\ee
for $k=6$. Note that the term $g^4\ln g$ has to be ascribed to the 
sixth--order approximation since only then one gets the series factoring 
$\ln g$, which could be summed.

The renormalization scale $\mu$ can be considered as a control function 
which can be defined from a fixed--point condition, e.g. from the 
minimal--difference condition [23]
\be
f_k(g,\mu) - f_{k-1}(g,\mu) = 0 \; ,
\ee
applied to the first $k$ approximation when the scale $\mu$ appears. In 
our case, this is $k=4$, for which condition (7) is equivalent to $a_4=0$ 
or $c_4=0$. Then we find
\be
\ln\frac{\mu}{T} \equiv \ln\gamma = 
\frac{0.04331\left ( 1 +\frac{1}{6}n_f\right )\ln\left ( 1 +\frac{1}{6}n_f
\right ) + 0.01733 -0.00763n_f -0.00088n_f^2}
{0.01323\left ( 1 +\frac{5}{12}n_f\right )\left ( 1 - \frac{2}{33}n_f
\right )} \; .
\ee
This gives $\mu\sim T$, which is physically reasonable since $T$ is the 
natural typical scale for the quark--gluon plasma. For example, 
$\gamma=3.70580$ if $n_f=0;\; \gamma=1.69564$ is $n_f=3$, and 
$\gamma=0.99696$ if $n_f=6$. Note that if one would take $k=5$ in 
condition (7), that is $a_4=c_4=0$, one would get a physically 
unreasonable value $\mu\sim 100 T$.

Following the method of self--similar exponential approximants [15,16], 
starting from the sequence $\{ f_k(g)\}$ with $f_k(g)$ given in (5), we
construct the exponentials
$$
F_2(g,\tau) = \exp\left ( b_2 g^2\tau\right ) \; ,
$$
\be
F_3(g,\tau) = 
\exp \left ( b_2 g^2\exp\left ( b_3 g\tau\right )\right ) \; ,
\ee
$$
F_5(g,\tau) =\exp\left ( b_2 g^2 \exp \left ( b_3 g\exp\left ( b_5 g^2\tau
\right )\right )\right ) \; ,
$$
in which $F_4=F_3$ because of (7), the coefficients are
$$
b_2 = a_2 = \frac{c_2}{c_0} \; , \qquad b_3 =\frac{a_3}{a_2} =
\frac{c_3}{c_2} \; , \qquad b_5 =\frac{a_5}{a_3} =\frac{c_5}{c_3} \; ,
$$ and $\tau=\tau_k(g)$ is a control function for the corresponding 
expression $F_k(g,\tau)$, so that
$$
F_k(g,\tau) - F_{k-1}(g,\tau) = 0 \; , \qquad
\tau =\tau_k(g) \; ,
$$ similarly to (7), with $\tau_2(g) \equiv 1$.

After defining the control functions $\tau_k(g)$, we obtain the self--similar
approximants
\be
f_k^*(g) \equiv F_k(g,\tau_k(g)) \; .
\ee
In particular,
$$
f_2^*(g) =\exp(b_2g^2) \; ,
$$
\be
f_3^*(g) =\exp (b_2g^2\tau_3(g)) \; ,
\ee
$$
f_5^*(g) =\exp\left ( b_2 g^2\exp\left ( b_3 g\tau_5(g)\right )\right )\; ,
$$
with the control functions defined by the equations
$$
\tau_3 =\exp (b_3 g\tau_3) \; , \qquad
\tau_5 = \exp ( b_5 g^2 \tau_5 ) \; .
$$

In expansion (4) and, respectively, in approximants (11), the coupling 
parameter $g=g(\mu)$ is the running coupling described by the 
renormalization group equation
\be
\mu \frac{\partial g}{\partial\mu} = \beta(g) \; .
\ee
The renormalization function $\beta(g)$, as $g\rightarrow 0$, has the 
expansion [24]
\be
\beta(g) \simeq \beta_3 g^3 + \beta_5 g^5 + \beta_7 g^7 \; ,
\ee
in which
$$
\beta_3 = -\frac{1}{(4\pi)^2} \left ( 11 - \frac{2}{3} n_f\right ) \; ,
\qquad \beta_5 = -\frac{2}{(4\pi)^4}\left ( 
51 - \frac{19}{3} n_f\right ) \; ,
$$
$$
\beta_7 = - \frac{1}{2(4\pi)^6} \left ( 2857 - \frac{5033}{9} n_f +
\frac{325}{27} n_f^2 \right ) \; .
$$
As the self--similar approximants for (13), we have
$$
\beta_5^*(g) = ag^3 \exp(bg^2) \; ,
$$
\be
\beta_7^*(g) = ag^3\exp\left ( bg^2\tau_7(g)\right ) \; ,
\ee
where the control function $\tau_7(g)$ is the solution of the equation 
$\tau_7=\exp(cg^2\tau_7)$ and the coefficients are $a\equiv\beta_3,\;
b\equiv \beta_5/\beta_3$, and $c\equiv \beta_7/\beta_5$.

To solve equation (12) with the right--hand side given by one of the 
functions from (14), we need an initial condition. For the latter, we may 
accept the value $\alpha_s(m_Z)=0.119$ of the strong coupling constant 
$\alpha_s = g^2/4\pi$ at the $Z^0$ boson mass $m_Z$ [25]. Thus, the 
initial condition for equation (12) is
$$
g(m_Z) = 1.222856\; , \qquad m_Z=91.187\; GeV\; .
$$
Solving (12), we find $g=g(\mu)$. Substituting the latter in (10) and 
using the relation $\mu=\gamma T$ following from (8), we obtain the 
reduced pressure
\be
\bar f_k(T) \equiv f_k^*(g(\gamma T))
\ee
as a function of temperature.

\section{Results and Conclusion}

The results of our calculations for $n_f=6$ are presented in Figs. 1 to 
3. The self--similar approximants $f_k^*(g)$ given by (11) for the 
reduced pressure (3), as functions of the coupling parameter $g$, are 
shown in Fig. 1. The solution $g(\mu)$ to the evolution equation (12) 
is drawn in Fig. 2. Both functions in (14) give practically the same 
solution $g(\mu)$. Figure 3 presents the equation of state (15).

\begin{figure}[h]

\psfig{figure=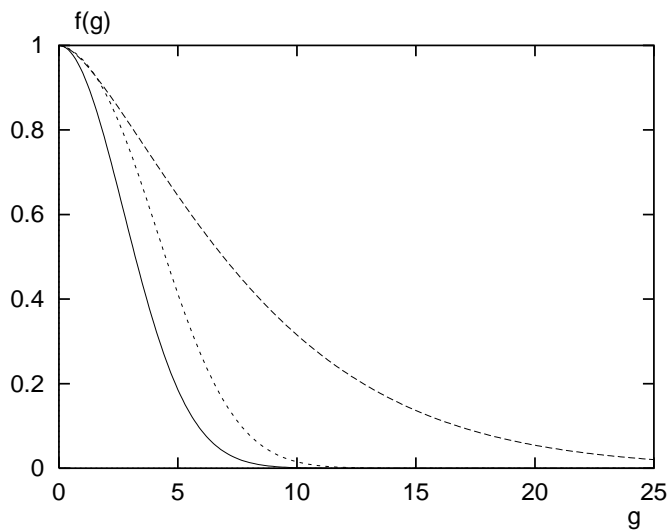,width=10cm,angle=270}
\caption{The behaviour of the reduced pressure $f(g)\equiv p(g)/p(0)$ 
presented by the self--similar approximations $f_2^*(g)$ (solid line), 
$f_3^*(g)$ (long--dashed line), and $f_5^*(g)$ (short--dashed 
line).}
\label{fig1}
\end{figure}

\vskip 5mm

\begin{figure}

\psfig{figure=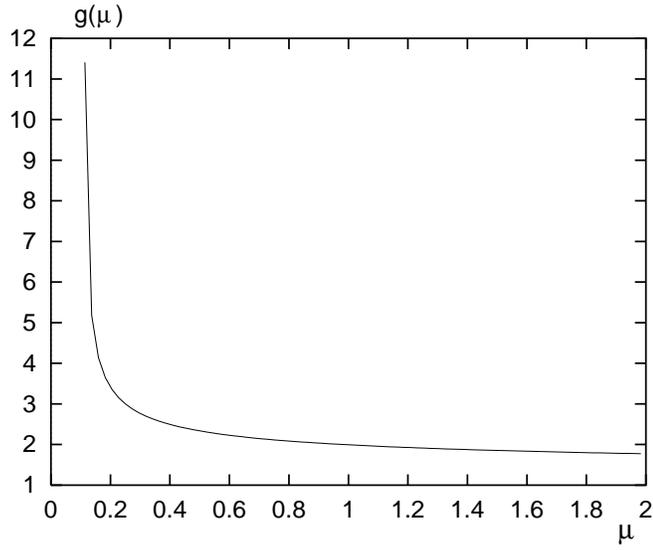,width=10cm,angle=270}
\caption{The running coupling $g(\mu)$ as a function of the scale $\mu$, 
obtained from the renormalization--group equation (12).}
\label{fig2}
\end{figure}

\vskip 5mm

\begin{figure}

\psfig{figure=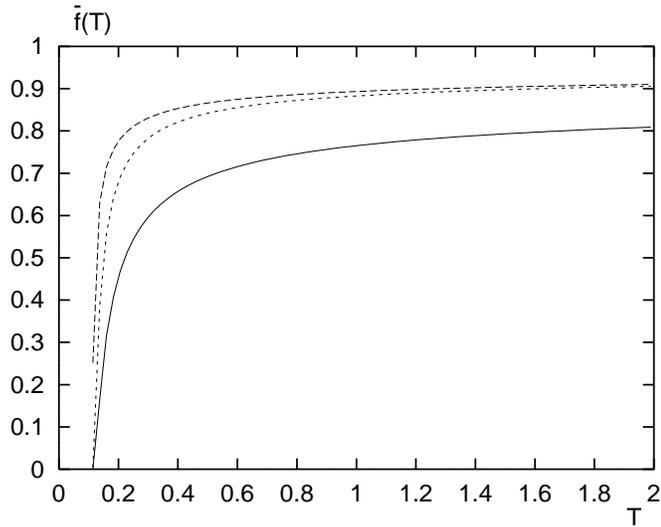,width=10cm,angle=270}
\caption{The reduced pressure $\bar f(T)$ given by the self--similar 
approximants $\bar f_2(T)$ (solid line), $\bar f_3(T)$ (long--dashed line), 
and $\bar f_5(T)$ (short--dashed line).}
\label{fig13}
\end{figure}

The overall behaviour of the reduced pressure $\bar f(T)$ is in agreement 
with that found in lattice simulations and in statistical modelling (see 
review [6]). At the temperature $T_c\approx (150-200)$MeV the pressure 
sharply drops down to practically zero, which can be interpreted as 
confinement. The curves corresponding to subsequent approximations 
$k=2,3,5$ are close to each other. The qualitative behaviour of $\bar 
f(T)$ for different numbers of flavours in the diapason $0\leq n_f\leq 8$ 
is the same, with increasing $T_c$ as $n_f$ decreases. Certainly, in the 
vicinity of $T_c\sim 200$MeV one should take into account the quark 
masses, especially those of heavy quarks. But what is very interesting is 
that, starting from perturbative expansions in the high--temperature region
of asymptotic freedom, we obtained an equation of state with the 
qualitatively correct behaviour, at all temperatures, including the 
existence of confinement at $T_c\sim 200$MeV.

\end{document}